\documentclass{Interspeech}

\interspeechcameraready

\usepackage{cite}
\usepackage{amsmath,amssymb,amsfonts}
\usepackage{algorithmic}
\usepackage{graphicx}
\usepackage{textcomp}
\usepackage{xcolor}

\usepackage{url}            
\usepackage{booktabs}       
\usepackage{amsfonts}       
\usepackage{xcolor}         
\usepackage{amssymb}
\usepackage{makecell}

\newcommand{\ie}{\textit{i}.\textit{e}.}
\newcommand{\eg}{\textit{e}.\textit{g}.}

\usepackage{pifont}
\newcommand{\cmark}{\ding{51}}
\newcommand{\xmark}{\ding{55}}
\usepackage{arydshln}
\usepackage{multirow}
\newcommand{\pmr}[1]{\scriptsize$\pm$#1}

\newcommand\blfootnote[1]{%
  \begingroup
  \renewcommand\thefootnote{}\footnote{#1}%
  \addtocounter{footnote}{-1}%
  \endgroup
}

\title{Revival with Voice: Multi-modal Controllable Text-to-Speech Synthesis}

\author[affiliation={1}]{Minsu}{Kim}
\author[affiliation={1,2}]{Pingchuan}{Ma}
\author[affiliation={1}]{Honglie}{Chen}
\author[affiliation={2}]{Stavros}{Petridis}
\author[affiliation={2}]{Maja}{Pantic}

\affiliation{}{Meta AI}{UK}
\affiliation{}{Imperial College London}{UK}
\email{ }

\usepackage{comment}

\begin{document}

\maketitle
\begin{abstract}
    This paper explores multi-modal controllable Text-to-Speech Synthesis (TTS) where the voice can be generated from face image, and the characteristics of output speech (\eg, pace, noise level, distance, tone, place) can be controllable with natural text description. Specifically, we aim to mitigate the following three challenges in face-driven TTS systems. 1) To overcome the limited audio quality of audio-visual speech corpora, we propose a training method that additionally utilizes high-quality audio-only speech corpora. 2) To generate voices not only from real human faces but also from artistic portraits, we propose augmenting the input face image with stylization. 3) To consider one-to-many possibilities in face-to-voice mapping and ensure consistent voice generation at the same time, we propose to first employ sampling-based decoding and then use prompting with generated speech samples. Experimental results validate the proposed model's effectiveness in face-driven voice synthesis.
\end{abstract}

\vspace{-0.1cm}
\section{Introduction}
\vspace{-0.1cm}
\blfootnote{Samples are  available: \url{https://bit.ly/4k5VU6b}. 

All experiments, data collection, and processing activities were conducted by ICL. Meta was involved solely in an advisory role and no experiments, data collection, or processing activities were conducted on Meta infrastructure.}
Text-to-Speech Synthesis (TTS) is one of the widely used technologies in diverse real-world applications~\cite{shen2018natural,elias2021parallel,renfastspeech,guo2023prompttts,kim2024textless,tan2024naturalspeech}. With recent developments, synthesized audio has become increasingly realistic and natural, and even the voice of outputs can be cloned with a few seconds of sample voices~\cite{wu2022adaspeech,wang2023neural}.

\begin{figure}[t]
    \begin{minipage}[b]{1.0\linewidth}  
        \centering
        \centerline{\includegraphics[width=8.0cm]{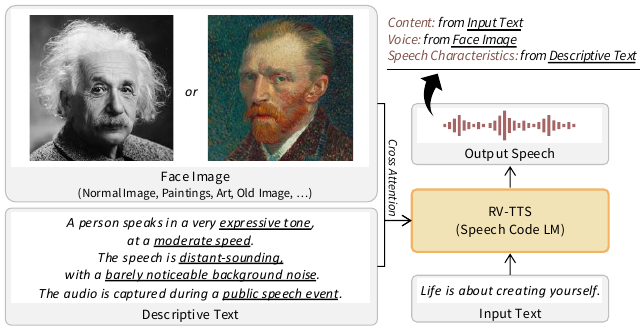}}
    \end{minipage}
    \vspace{-0.4cm}
    \caption{Overview of the proposed RV-TTS: The face image controls the voice, the descriptive text controls speech characteristics, and the input text determines the content of speech.}
    \label{fig:1}
    \vspace{-0.6cm}
\end{figure}

In this paper, we explore an interesting extension of TTS, where the voice is generated to match a given face image, even including artistic portraits. This task can be applied to synthesize the voice of past individuals or non-existent characters who are only known through remaining artworks, old photos, or other visual records (\eg, Imagine the Mona Lisa explaining the painting in her own voice, bringing a new level of depth and intimacy to one of the world's most famous works of art). Going one step forward, we try to put controllability of the speech characteristics (\ie, pace, noise-level, distance to mic, tone, and the places of the recording) with natural text description. Therefore, as illustrated in Fig.~\ref{fig:1}, the synthesized speech voice is controlled by the input face image, while other speech characteristics are determined by the given descriptive text, and the content of the speech is controlled by the input text as usual.

To train these face-driven TTS models~\cite{goto2020face2speech,park2024synthe,guan2024mm}, we have to employ audio-visual speech data where face and its speech are paired (\eg, LRS2~\cite{chung2017lip} and LRS3~\cite{afouras2018lrs3}). However, most publicly available large-scale audio-visual speech datasets are captured in public events, which puts challenges in developing a high-quality face-driven TTS model; 1) The data often contains noisy audio and includes stuttering, whereas clean TTS data~\cite{koizumi2023librittsr,pratap2020mls} is essential for high-quality TTS training. 2) The data is based on real humans, which means that the model trained with this data may exhibit degraded performance when tested with a face from an artwork. 3) Face-to-voice mapping is ambiguous, as multiple voices can be imagined from a single face. These challenges have led previous works~\cite{lee2023imaginary,lee2024fvtts} to generate public speech-like audio with noticeable background noise, to have limited applicability to diverse face images, and to generate inconsistent voice for the same speaker.

To mitigate these challenges, we adopt three main strategies. Firstly, to improve speech quality, we propose employing not only large-scale audio-visual speech datasets but also high-quality audio-only speech datasets in model training through alternating voice embeddings. To this end, the face encoder and the audio encoder are pre-trained using a contrastive learning objective, such that the resulting face-driven voice embedding and audio-driven voice embedding, which are embedded through each encoder, can share the same latent space. Then, the proposed model is conditioned on either one of the voice embedding according to the sampled type of data; \ie, if the data is from audio-visual speech dataset, use face-driven voice embedding, otherwise use audio-driven voice embedding. Secondly, to bridge the gap between artistic portraits and real human, we propose augmenting the input face image with a pre-trained style transfer~\cite{gatys2016image} where the input face image is randomly stylized with different style images. Finally, to reflect the multiple nature of face-to-voice mapping, we employ sampling-based decoding to obtain multiple voices. Subsequently, to produce a consistent voice, we use the generated sample as a prompt for our speech code language model.

\begin{figure*}[t]
    \begin{minipage}[b]{1.0\linewidth}  
        \centering
        \centerline{\includegraphics[width=17.0cm]{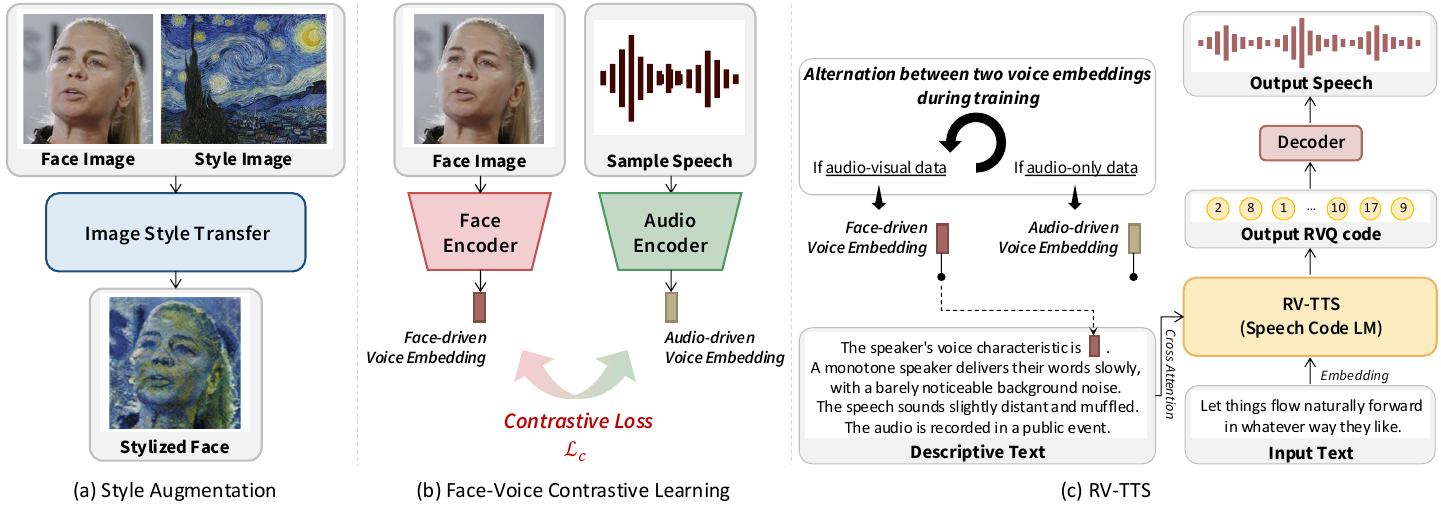}}
    \end{minipage}
    \vspace{-0.7cm}
    \caption{Illustration of the proposed RV-TTS. (a) The face image is randomly stylized using a pre-trained style transfer model to reduce the gap between real human faces and artistic portraits. (b) The face encoder and audio encoder are pre-trained through contrastive learning to share a common embedding space. (c) During training, the model alternates between face-driven and audio-driven voice embeddings to learn not only to associate face images with voices but also to synthesize high-quality audio.}
    \label{fig:2}
    \vspace{-0.6cm}
\end{figure*}

Based on the above strategies, we propose RV-TTS (\textbf{R}evival with \textbf{V}oice). It takes three inputs, a face image, natural descriptive text, and input text. The face image is encoded into a face-driven voice embedding by a face encoder. The natural descriptive text is embedded through a pre-trained language model~\cite{raffel2020t5}. Then, the descriptive text embedding and the voice embedding are concatenated, and used to condition the generation process through cross-attention. RV-TTS employs Residual Vector Quantized (RVQ) code~\cite{kumar2024rvq} as the speech features and learns speech code language modeling.

Our contribution can be summarized as follows: 1) We propose a multi-modal controllable text-to-speech synthesis model, where the voice of output speech can be inferred from a facial image and the speech characteristics can be controlled with natural descriptive text. 2) To overcome the limited speech quality of current audio-visual speech corpora, we enforce training by mixing both audio-visual and high-quality audio-only databases by alternating the voice embeddings of the two modalities. 3) To generate the voices of past individuals whose voice recordings are no longer available, but whose faces have been preserved through paintings or old photographs, we propose to augment our training with stylization. 4) The proposed RV-TTS can imagine multiple voices matched to the given face image, while it can generate consistent voice through prompting. 

\vspace{-0.1cm}
\section{Method}
\vspace{-0.1cm}
Let $x$ be input text which will be read by the TTS model, $y$ be the ground-truth speech, $c_i$ be the face image, $c_a$ be the sample utterance, and $c_t$ be the descriptive text. The objective of our learning problem is to synthesize speech $y$ that reads the input text $x$ with the voice characteristics of the conditioned face image $c_i$, while also capturing the other speech attributes (\ie, pace, tone, noise level, distance, and place) described in $c_t$.

\vspace{-0.1cm}
\subsection{Architecture and Overview}
\vspace{-0.1cm}
The proposed TTS system follows~\cite{lyth2024parlertts} whose architecture is based on that of MusicGen~\cite{copet2024simple}. It is a speech language model, constructed using a Transformer architecture with cross-attention layers~\cite{vaswani2017attention}, that auto-regressively predicts 9-level RVQ codes of~\cite{kumar2024rvq}. Each higher level code is shifted to the right, forming the delay pattern. The architecture of the face encoder is ResNet50~\cite{he2016deep} and that of audio encoder is ECAPA-TDNN~\cite{desplanques2020ecapa}. As in \cite{lyth2024parlertts}, a pre-trained language model, T5~\cite{raffel2020t5}, is used to encode the natural descriptive text, $c_t$. The overall flow of the proposed RV-TTS is shown in Fig.~\ref{fig:2}c. The input text $x$ is embedded through an embedding layer and passed into the speech language model as inputs. The face image $c_i$ is encoded into a face-driven voice embedding using the face encoder. The descriptive text $c_t$ is embedded through a pre-trained language model. Then, the descriptive text embedding and the face-driven voice embedding are concatenated along the temporal dimension and passed into the cross-attention layer of the speech language model. By conditioning on all inputs, the speech language model predicts RVQ codes, which are then converted into a raw waveform through decoder of RVQ model.

\vspace{-0.1cm}
\subsection{Style Augmentation}
\vspace{-0.1cm}
Paired face image and speech data is crucial for training a face-driven TTS system. However, collecting such paired data for historical figures, particularly those whose likenesses are preserved only in visual records, is impossible. Therefore, previous literature has typically utilized publicly available audio-visual speech data, which consists of recordings of real-world individuals. Since the model is trained on photorealistic face images, this creates a discrepancy between training and inference when the model is applied to non-realistic faces (\eg, portraits), which limits its performance and applicability to diverse faces.

To mitigate this discrepancy, we augment the face image during training with diverse styles by employing neural style transfer models~\cite{gatys2016image}, as shown in Fig.~\ref{fig:2}a. Specifically, the input face image is randomly passed through a style transfer model, CAST~\cite{zhang2022cast}, and stylized with a random style image (\eg, paintings and sketches). Additionally, to simulate old photographs, the images are also augmented with gray-scaling and blurring.

\vspace{-0.1cm}
\subsection{Narrowing the Speech Quality Gap}
\vspace{-0.1cm}
Another consideration is that the audio quality of publicly available large-scale audio-visual datasets is often compromised by diverse background noise and reverberation. Consequently, this limits the face-driven TTS system's ability to learn high-quality audio synthesis, unlike typical TTS systems that are usually trained on high-quality datasets captured from reading speech \cite{zen2019libritts}. To address this limitation, our strategy is to also employ a high-quality audio-only dataset, LibriTTS-R~\cite{koizumi2023librittsr,koizumi2023miipher}, in the training of the face-driven TTS model.

To this end, we first pre-train the face encoder and the audio encoder to share their embeddings through contrastive learning~\cite{oord2018representation} as shown in Fig.~\ref{fig:2}b. Specifically, the face-driven voice embedding $z_i$ is obtained using the face encoder from face image $c_i$ and similarly the audio-driven voice embedding $z_a$ from the sample utterance $c_a$ using the audio encoder. Then, if the $z_i$ and $z_a$ come from the same person, we maximize their similarity, otherwise minimize using InfoNCE loss \cite{oord2018representation,chen2020simple} as follows:
\begin{equation}
\setlength{\abovedisplayskip}{3pt}
\setlength{\belowdisplayskip}{4pt}
    \mathcal{L}_{c} = - \frac{1}{N} \sum_{l=1}^N \log \frac{\exp(s(z_i^l, z_a^l)/\tau)}{\sum_{n=1}^N\exp(s(z_i^l, z_a^n)/\tau)},
\end{equation}
where the superscript is the speaker ID, $\tau$ is a learnable temperature parameter, $s(\cdot)$ denotes the cosine similarity function, and $N$ represents the total number of speakers in a mini batch. 

After pre-training the face and voice encoders through contrastive learning, we can regard the two voice embeddings obtained from each modality as sharing similar information. This means that we can use either one of the embeddings as a condition to generate the target voice. Concretely, when the data is sampled from an audio-visual dataset, the face-driven voice embedding $z_i$ is passed to the TTS model, while $z_a$ is passed if the data comes from an audio-only dataset, as shown in Fig. \ref{fig:2}c. By alternatively training the model with both modalities, the model can learn not only to associate the target voice with a given face image but also produce high-quality speech (\ie, noise-clean and stuttering-free).

\vspace{-0.2cm}
\subsection{Diverse but Consistent Voice Generation}
\vspace{-0.1cm}
Inferring the voice by just watching a face is an ill-posed problem, as multiple voices can potentially match a single face. However, previous methods have not explored this one-to-many possibilities and attempted to produce a single fixed voice for a given face image. Unfortunately, it often results in inconsistent voices across multiple inferences, when using different input text even with the same face image.

However, this limitation can be easily addressed in the proposed system. Since the proposed RV-TTS is based on a speech language model, we can generate different voices that match the input face through sampling-based decoding. Furthermore, we can also generate consistent voices across multiple inferences by providing the previously generated sample as a prompt, which is also known as in-context learning ability~\cite{brown2020language,wang2023neural}. Therefore, users can employ RV-TTS to produce diverse sentences with their preferred voice using the following steps: 1) Generate multiple sample speeches with the proposed system by providing a simple input text (\eg, `voice candidate') and the target face image. 2) Choose one plausible sample speech based on the user's preference. 3) Generate target speech with the proposed system by providing the selected sample speech at the previous step, the target input text, and the target face image. That is, by employing the selected sample speech as prompt, the model can produce a consistent voice.

\vspace{-0.2cm}
\subsection{Natural Descriptive Text}
\vspace{-0.1cm}
To make RV-TTS be able to control speech characteristics using natural descriptive text, descriptive labels are required for each speech. For the natural descriptive text labels of LibriTTS-R, we can employ the labels released by~\cite{lyth2024parlertts}. However, these texts contain gender-specific words (\eg, She, He, Her, His), which might impede the model's ability to learn to associate voice with face images only. To resolve this issue, we remove gender-specific words and replace them with neutral words (\eg, A speaker, A person) before training. For LRS3 and VoxCeleb2, since there are no natural descriptive texts available, we generate labels using Data-Speech~\cite{lacombe-etal-2024-dataspeech}, which analyzes diverse audio characteristics (\eg, SNR, pitch, speaking rate) and generates text. Moreover, we added a sentence describing the audio as being recorded in public speech events to inform the data is from broadcasts or TED talks. Finally, to incorporate the face- or audio-driven voice embedding with the natural descriptive text embedding, we insert the voice embedding between the embeddings of `The speaker's voice characteristic is' and the remaining natural descriptive text, as shown in Fig.~\ref{fig:2}c.

\vspace{-0.2cm}
\section{Experiments}
\vspace{-0.2cm}
\subsection{Dataset}
\vspace{-0.1cm}
We employ audio-visual speech dataset, LRS3 and VoxCeleb2, and audio-only high-quality speech dataset, LibriTTS-R, to train RV-TTS. For evaluation, we employ the test set of LRS3 and artistic portraits, a collection of 20 copyright-free images.

\textbf{LRS3}~\cite{afouras2018lrs3} consists of over 400 hours of video collected from TED and TEDx talks. It contains face video, speech, and transcription. Evaluation is performed on the test set of LRS3.

\textbf{VoxCeleb2}~\cite{chung2018voxceleb2} is an unlabeled audio-visual speech data collected from YouTube. We employ the English video only following~\cite{shi2022learning} and generate the transcription by using a pre-trained Automated Speech Recognition (ASR) model~\cite{radford2023robust} following~\cite{ma2023auto}. The resulting data is about 1,300 hours.

\textbf{LibriTTS-R}~\cite{koizumi2023librittsr} is an audio-only TTS dataset which is the sound quality improved version of LibriTTS~\cite{zen2019libritts} by applying speech restoration~\cite{koizumi2023miipher}. It contains 585 hours of audio.

\vspace{-0.2cm}
\subsection{Implementation Details}
\vspace{-0.1cm}
RV-TTS is composed of 24 Transformer layers with 16 heads, 1,024 hidden dimension, 4,096 feed-forward dimension. The codebook size of RVQ code~\cite{kumar2024rvq} is 1,024 with 9 levels. The face encoder is initialized with a pre-trained ArcFace~\cite{deng2019arcface} ResNet50 model, and the audio encoder is initialized with a pre-trained ECAPA-TDNN~\cite{desplanques2020ecapa} model whose input is modified with RVQ codes instead of MFCCs. The face image is preprocessed according to~\cite{deng2019arcface}. During the contrastive pre-training, a linear layer with a hidden dimension of 256 is attached for both face and audio encoders, and audio encoder is frozen except the attached linear layer. From the video, one face image is randomly selected, and a random segment of audio between 3 to 5 seconds is chosen. The contrastive pre-training is performed with a batch size of 256, an initial learning rate of $5e^{-5}$, cosine learning rate scheduler, and AdamW optimizer \cite{kingma2015adam,loshchilov2019adamw} for 30k steps on VoxCeleb2 dataset. For training RV-TTS, a linear layer is employed to embed the voice embedding of 256 dimension into 1,024. It is trained with a batch size of 24 and an initial learning rate of $5e^{-4}$ for 500k steps on LRS3, VoxCeleb2, and LibriTTS-R datasets. For style augmentation, the image is augmented with a 50\% chance among one of the following: style transferring, gray-scaling, and blurring, each with a uniform probability. For style images, copyright-free paintings, photos, and artworks are used. For decoding, top 30 sampling is performed with a repetition penalty of 1.2.

\vspace{-0.2cm}
\subsection{Metrics}
\vspace{-0.1cm}
To evaluate the performance of the proposed face-driven TTS system, we conduct human subjective evaluations as objective metrics often do not accurately reflect perception~\cite{wang2017tacotron}. We collect Mean Opinion Scores (MOS) to assess the naturalness of the synthesized speech, Face Matching Scores (FMS) to evaluate how well the predicted speech matches the input face image, and Voice Consistency Scores (VCS) to measure the consistency of the voice across different predicted speeches with the same face image. The MOS and FMS scores were measured from 20 samples of LRS3, while the VCS score was measured from 3 samples for each of 10 speakers (\ie, 30 samples), for each method, unless otherwise specified.

\subsection{Experimental Results}
\vspace{-0.1cm}
\subsubsection{Ablation study on each proposed component}
\vspace{-0.1cm}
Results for the ablation study are shown in Table~\ref{table:1}. We remove one component at a time to estimate its contribution to the final model. In particular, we observe an absolute drop of 0.5 and 0.29 in both MOS and FMS scores, respectively, when High-Quality (HQ) audio data (\ie, LibriTTS-R) are removed, which indicates that using clear audio data is the most important component to improve the performance of the face-driven TTS model. The performance further drops by 0.26 and 0.16 in MOS and FMS by removing style augmentation, which suggests that training with style augmentation can provide not only better face matching but also improved naturalness in the generated speech. However, when replacing contrastive learning with a pre-trained encoder~\cite{deng2019arcface} to extract face embeddings, we observe the FMS score is reduced while MOS score slightly improved. These results are in line with the hypothesis that the use of contrastive learning are primarily contributes to preserving face-voice association. To further confirm this, we additionally measure the voice similarity (SIM) between the predicted speech and the ground-truth speech by employing a pre-trained speaker recognition model~\cite{wang2024advancing}. The results show that SIM degrades if contrastive learning is not performed, indicating a loss of face-voice association ability.

\begin{table}[t]
    \renewcommand{\arraystretch}{1.3}
    \renewcommand{\tabcolsep}{2mm}
    \caption{Ablation study on each proposed component. The MOS and FMS scores were measured from 30 samples of both LRS3 and artistic portraits for each method.}
    \centering
\vspace{-0.3cm}
\resizebox{0.99\linewidth}{!}{
\begin{tabular}{ccc ccc}
\Xhline{3\arrayrulewidth}
\multicolumn{3}{c}{\textbf{Method}} & \multirow{2}{*}{\textbf{MOS}} & \multirow{2}{*}{\textbf{FMS}} & \multirow{2}{*}{\textbf{SIM}} \\ \cline{1-3}
\textbf{HQ audio} & \textbf{Style Aug.} & \textbf{Contrastive} & & & \\ \hline
\cmark & \cmark & \cmark & 4.11\pmr{0.18} & 3.73\pmr{0.32} & 0.11 \\ \hdashline
\xmark & \cmark & \cmark & 3.61\pmr{0.21} & 3.44\pmr{0.24} & 0.11 \\ 
\xmark & \xmark & \cmark & 3.35\pmr{0.20} & 3.28\pmr{0.20} & 0.11 \\ 
\xmark & \cmark & \xmark & 3.73\pmr{0.22} & 3.29\pmr{0.27} & 0.09 \\ 
\Xhline{3\arrayrulewidth}
\end{tabular}}
\vspace{-0.2cm}
\label{table:1}
\end{table}
\begin{table}[t]
    \renewcommand{\arraystretch}{1.3}
    \renewcommand{\tabcolsep}{3mm}
    \caption{Human subjective evaluation score comparisons.}
    \centering
\vspace{-0.3cm}
\resizebox{0.99\linewidth}{!}{
\begin{tabular}{l cccc}
\Xhline{3\arrayrulewidth}
\textbf{Method\quad\quad\quad\quad\quad\quad\quad\quad} & \textbf{MOS} & \textbf{FMS} & \textbf{VCS} \\ \hline
\multicolumn{3}{l}{$\bullet$ \textit{Random Voice Synthesis}} \\
\quad Parler-TTS~\cite{lyth2024parlertts} & 3.68\pmr{0.34} & 1.88\pmr{0.18} & - \\ \hline
\multicolumn{3}{l}{$\bullet$ \textit{Audio-driven Voice Synthesis}} \\
\quad YourTTS~\cite{casanova2022yourtts} & 2.78\pmr{0.27} & 3.09\pmr{0.33} & \textbf{4.42}\pmr{0.35} \\ \hline
\multicolumn{3}{l}{$\bullet$ \textit{Face-driven Voice Synthesis}} \\
\quad FaceTTS~\cite{lee2023imaginary} & 1.84\pmr{0.36} & 2.31\pmr{0.27} & 2.85\pmr{0.47} \\
\quad FVTTS~\cite{lee2024fvtts} & 1.60\pmr{0.25} & 2.38\pmr{0.25} & 3.45\pmr{0.52} \\
\quad \textbf{RV-TTS} & \textbf{4.14}\pmr{0.20} & \textbf{3.86}\pmr{0.35} & 3.96\pmr{0.19} \\
\Xhline{3\arrayrulewidth}
\end{tabular}}
\label{table:2}
\vspace{-0.6cm}
\end{table}

\vspace{-0.1cm}
\subsubsection{Comparisons with previous methods}
\vspace{-0.1cm}
We compare our proposed RV-TTS with previous methods. In particular, we employ Parler-TTS~\cite{lyth2024parlertts}, a random voice synthesis model that depends on the speaker's name in the text description; YourTTS~\cite{casanova2022yourtts}, an audio-driven voice synthesis model; and FaceTTS~\cite{lee2023imaginary} and FVTTS~\cite{lee2024fvtts}, face-driven voice synthesis models. Results on LRS3 and artistic portraits are shown in Table~\ref{table:2}. We see that overall the proposed RV-TTS model outperforms other works in both MOS and FMS scores, respectively, which highlights that the proposed system can generate high-quality audio with voices that are consistent and well-matched to the input face. When comparing to one of the best previous face-driven TTS methods, FaceTTS, we see that the proposed RV-TTS leads to a substantial improvement of 2.30 in MOS and 1.45 in FMS, respectively, indicating its high-quality speech synthesis capabilities. In terms of VCS, the audio-driven voice synthesis method, YourTTS, achieved the best score. Among face-driven TTS methods, the proposed RV-TTS outperforms previous methods. 

\begin{figure}[t]
    \begin{minipage}[b]{1.0\linewidth}  
        \centering
        \centerline{\includegraphics[width=8.0cm]{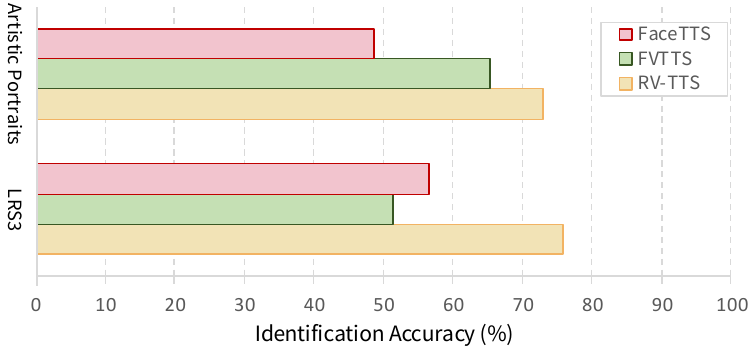}}
    \end{minipage}
    \vspace{-0.6cm}
    \caption{Speaker identification test results comparisons.}
    \label{fig:3}
    \vspace{-0.6cm}
\end{figure}

In order to better understand how the generated speech match the provided face image, we perform a speaker identification test, where participants are asked to select one face image out of two that best matches the generated speech. A total of 78 samples were rated by 15 people. Identification accuracy in both LRS3 and artistic portraits are shown in Fig.~\ref{fig:3}. It is clear that our proposed RV-TTS outperforms the two previous face-driven TTS methods and achieved the highest accuracies of 76\% and 73\% on LRS3 and artistic portraits, respectively. These results highlight that the proposed RV-TTS can produce better voices not only when the face image is photorealistic but also when it is an old photo or artwork.

\begin{table}[t]
    \renewcommand{\arraystretch}{1.5}
    \renewcommand{\tabcolsep}{0.1mm}
    \caption{Analysis on controllability using descriptive text.}
    \centering
\vspace{-0.3cm}
\resizebox{0.99\linewidth}{!}{
\begin{tabular}{c ccc}
\Xhline{3\arrayrulewidth}
\textbf{Feature} & \textbf{Low} & \textbf{Moderate} & \textbf{High} \\ \hline
\makecell{\textbf{Pace} \\ (Speaking Rate)} & \makecell{slow \\ 7.64} & \makecell{moderate \\ 12.42} & \makecell{fast \\ 17.70} \\ \hline
\makecell{\textbf{Noise} \\ (SI-SDR)} & \makecell{almost noiseless \\ 24.39} & \makecell{slight noise \\ 21.33} & \makecell{very noise \\ 19.37} \\ \hline
\makecell{\textbf{Distance} \\ (C50)} & \makecell{very close \\ 57.73} & \makecell{moderate distant \\ 49.94} & \makecell{very distant \\ 47.33} \\ \hline
\makecell{\textbf{Tone} \\ (Pitch Std)} & \makecell{monotone \\ 32.66} & - & \makecell{expressive \& animated \\ 91.46} \\ \hline
\makecell{\textbf{Place} \\ (C50 \& SNR)} & \makecell{not at public speech \\ 57.73 \& 51.57} & - & \makecell{at public speech \\ 52.79 \& 47.96} \\
\Xhline{3\arrayrulewidth}
\end{tabular}}
\vspace{-0.13cm}
\label{table:3}
\end{table}

\vspace{-0.1cm}
\subsubsection{Analysis on controllability using descriptive text}
\vspace{-0.1cm}
To evaluate the controllability of RV-TTS, we measured each factor of the controlled feature using corresponding metrics; Pace: Speaking Rate that is the number of phonemes divided by the duration of speech, Noise: Scale Invariant Signal Distortion Ratio (SI-SDR)~\cite{le2019sdr}, Distance: Clarity using C50~\cite{bradley1999just}, Tone: Standard deviation (Std) of pitch, Place (\ie, Public space or not): C50 and Signal to Noise Ratio (SNR). We estimate SI-SDR, SNR, and C50 by using pre-trained models~\cite{kumar2023torchaudio,lavechin2023brouhaha}, following \cite{lyth2024parlertts}. The analysis results on LRS3 test set are shown in Table~\ref{table:3}. We can confirm that changing the words in the natural descriptive text affects the corresponding metrics. For example, using the word `monotone' results in a mean Std of pitch of 32.66~Hz, while using the phrase `expressive and animated tone' results in a mean Std of pitch of 91.46~Hz.

\section{Conclusion}
We proposed RV-TTS, a multi-modal controllable TTS model. By employing style augmentation during training, the model can generate voices from artistic portraits as well as normal human face images. By alternating voice embeddings from different modalities, the model successfully learns to associate faces with voices and generates high-quality audio. The evaluation results confirm that RV-TTS outperforms previous face-driven TTS methods and offers controllability using descriptive text.

\clearpage
\bibliographystyle{IEEEtran}
\bibliography{mybib}

\end{document}